\documentclass[aps,prb,twocolumn,superscriptaddress]{revtex4}
\usepackage{graphicx}

\begin{document}

\title{Impurity effects in quantum dots: Towards quantitative modeling}

\author{E.~R\"as\"anen}
\email[Electronic address: ]{ehr@fyslab.hut.fi}
\affiliation{Laboratory of Physics, Helsinki University of Technology,
P.O. Box 1100, FIN-02015 HUT, Finland}
\author{J. K\"onemann}
\affiliation{Institut f\"ur Festk\"orperphysik, Universit\"at Hannover,
Appelstrasse 2, D-30167 Hannover, Germany}
\author{R. J. Haug}
\affiliation{Institut f\"ur Festk\"orperphysik, Universit\"at Hannover,
Appelstrasse 2, D-30167 Hannover, Germany}
\author{M.~J.~Puska}
\affiliation{Laboratory of Physics, Helsinki University of Technology,
P.O. Box 1100, FIN-02015 HUT, Finland}
\author{R.~M.~Nieminen}
\affiliation{Laboratory of Physics, Helsinki University of Technology,
P.O. Box 1100, FIN-02015 HUT, Finland}

\date{\today}

\begin{abstract}

We have studied the single-electron transport spectrum of a 
quantum dot in GaAs/AlGaAs resonant tunneling device. The measured 
spectrum has irregularities indicating a 
broken circular symmetry. We model the system with an external 
potential consisting of a parabolic confinement and a 
negatively charged Coulombic impurity placed in the vicinity of the 
quantum dot. 
The model leads to a good agreement between the calculated
single-electron eigenenergies and the experimental spectrum. 
Furthermore, we use the spin-density-functional theory to 
study the energies and angular momenta when the system contains 
many interacting electrons.
In the high magnetic field regime the increasing electron 
number is shown to reduce the distortion induced by the impurity.
\end{abstract}

\maketitle

\section{Introduction}

The tunability in size, shape, and electron 
number of semiconductor quantum dots (QD) 
provides numerous technological applications
as well as interesting many-electron physics.~\cite{revmod}
In actual QD devices, the effects induced
by impurities or donor scattering centers
may be remarkable. In most cases, irregularities 
in samples have only an indirect influence 
on the many-body structure, complicating 
the identification of the origin behind 
the peculiar behavior in the measured 
characteristics of QD's.

A clean quantum dot typically shows
single-electron energy levels reminiscent
of the well-known Fock-Darwin (FD) energy spectrum 
corresponding to a parabolic confining 
potential.~\cite{fd} 
Adding external impurities into the QD breaks 
the circular symmetry of the system,
leading to avoided crossings and liftings of
the degeneracies in the single-electron 
energy spectrum. 
This was demonstrated by 
Halonen~{\em et al.},~\cite{halonen}
who studied theoretically QD's distorted by 
repulsive Gaussian scattering 
centers. However, even if clear traces of the FD spectrum have 
been obtained experimentally in both lateral~\cite{mceuen}
and vertical~\cite{vertical,tarucha} quantum dots, there is, to the
best of our knowledge, 
no direct experimental evidence of repulsive impurities 
present in QD structures. 
Instead, states bound to hydrogenic impurities, probably 
arising from Si dopant atoms in the GaAs quantum well,
were found already by Ashoori and 
co-workers~\cite{ashoori} in their pioneering 
single-electron tunneling experiment. 
These impurities have been
suggested to be sources of pair-tunneling states,
theoretically analyzed with a superimposed
attractive $1/r$-type potential.~\cite{wan,lee}

Theoretically, the distortion of the circular 
symmetry makes the many-electron problem 
particularly complex to solve, especially in the
presence of an external magnetic field.
In the above-mentioned study, 
Halonen~{\em et al.}~\cite{halonen} applied
exact diagonalization up to three electrons
and focused on the effects of impurities
on the energy levels and optical absorption
spectra. 
Recently, G\"u\c{c}l\"u and co-workers~\cite{gycly} 
performed diffusion quantum
Monte Carlo (DMC) calculations on QD's distorted by
randomly distributed Gaussian scatterers and
studied the energetics up to ten electrons. 
They found that in these systems the transitions
between the many-body states are 
considerably less pronounced than in clean
dots. Hirose and Wingreen~\cite{hirose} have
used the spin-density-functional theory (SDFT)
to examine the energies and spin states in 
disordered QD's as a function of the interaction 
strength in zero magnetic field.
Besides additional scatterers, non-circular
QD's have attracted general interest in
connection with the chaotic properties~\cite{ahn} or
the behavior in the high magnetic field limit.~\cite{recta}

In this paper we present a measured single-electron 
transport spectrum where avoided crossings and
lifted degeneracies are clearly observable.
We reproduce the spectrum with an appropriate
model potential, showing that the unexpected 
effects in the spectrum result from a
negatively charged Coulombic impurity located
near the QD.
The many-electron properties studied by the SDFT
reflect the strongly distorted single-electron
spectrum. The variation of the impurity location
shows the stability of the maximum-density droplet (MDD)
and the screening of the impurity by electrons.

The outline of this paper is as follows. 
In Sec.~\ref{sec1} we briefly describe the fabrication 
of the sample and report the transport measurement.
In Sec.~\ref{sec2} the theoretical model describing
the physical system is given and the single-electron calculations
are compared to the experiment. In Sec.~\ref{sec4} the
many-electron properties, i.e., 
chemical potentials, MDD stability, and total
magnetization are studied with the SDFT. 
The paper is summarized in Sec.~\ref{sec5}.


\section{Experiment} \label{sec1}

\begin{figure}
\includegraphics[width=6cm]{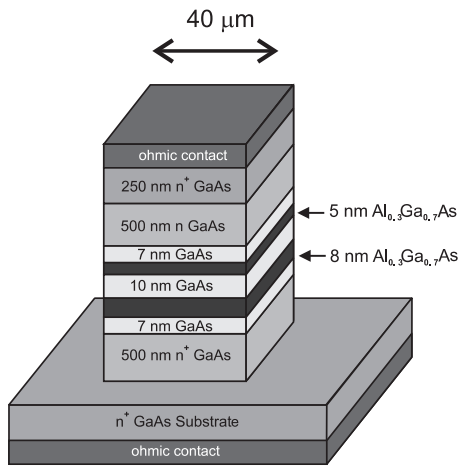}
\caption{Sketch of the heterostructure of our sample.} \label{sample}
\end{figure}

The heterostructure consists of a 10~nm wide GaAs quantum well sandwiched between two
Al$_{0.3}$Ga$_{0.7}$As-tunneling barriers of $5$ and $8$~nm, see Fig.~\ref{sample}.
The contacts are formed by $0.5$~$\mu$m thick GaAs layers highly doped with Si up to $4\times
10^{17}$~cm$^{-3}$ and separated from the active region by 7~nm thin spacer layers of
undoped GaAs. Our sample was defined as a mesa of 40~$\mu$m size.
We carried out direct-current measurements of the current-voltage 
(I-V) characteristics 
in a He$^3$-refrigerator at 350~mK base 
temperature in magnetic fields up to $14$~T.

\begin{figure}
\vspace{0.2cm}
\includegraphics[width=8.5cm]{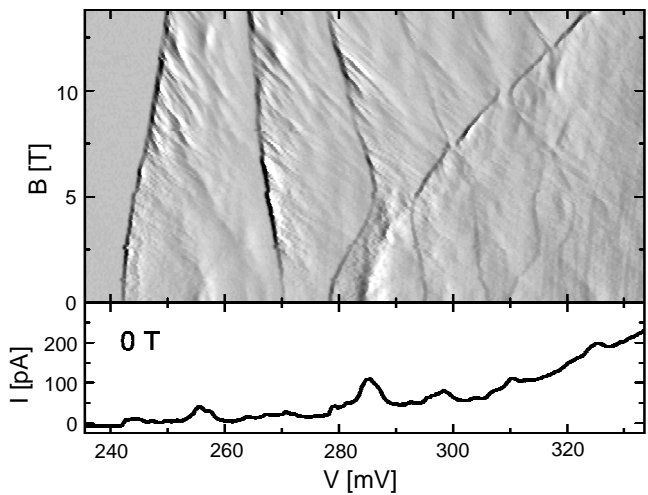}
\caption{Top: G(V,B) plot of the transport spectrum of our sample. Bottom: I-V
characteristics for $B=0$~T.} \label{exp_spectrum}
\end{figure}

Figure~\ref{exp_spectrum} 
shows the resulting transport spectrum of a quantum dot formed in a local
potential minimum. 
The black lines $V_{n,l}$ correspond to high differential conductance $G={\rm d}I/{\rm d}V$.
They trace the position of the single-electron energy states $E_{n,l}$ of the spectrum
according to a relation
\begin{equation} \label{conv}
V_{n,l}=V_0+1/(e\alpha) E_{n,l},
\end{equation}
where the energy-voltage conversion factor $\alpha$ equals 0.4, determined from measurements of
the broadening of the step edge with temperature, and the onset voltage is fitted to
be $V_0=172$~mV. The energy levels $V_{n,l}$ in the transport spectrum
can be interpreted as single-particle levels of a local, 
presumably a grown-induced potential minimum in the GaAs quantum well of our device. The
lowest FD band is clearly visible, and one can also observe higher excited states. In contrast 
to ordinary FD energy levels, we are able to observe broken
energy degeneracies at $B=0$~T and strong anticrossing effects in the spectrum.

\section{Modeling the quantum dot} \label{sec2}

We expect the quantum well confined in the GaAs layer to have
a negligible degree of freedom for electrons in the vertical direction.
Our model system is thus strictly two-dimensional and 
defined to be located on the $xy$ plane.
The single-electron Hamiltonian is written as
\begin{equation} \label{hami1}
h({\mathbf r})=\frac{1}{2m^*}\left[{\mathbf p}+e{\mathbf A}({\mathbf r})\right]^2
+V_{\rm conf}({\mathbf r})+V_{\rm imp}({\mathbf r}),
\end{equation}
where we use the effective-mass approximation with $m^*=0.067\,m_e$,
which is the typical value for electrons moving in GaAs.
In a symmetric gauge the vector potential reads
as $\mathbf{A}=\frac{B}{2}(-y,x,0)$, giving 
the external magnetic field ${\mathbf B}=B\hat{z}$
perpendicular to the QD plane. 
The Zeeman energy is omitted in Eq.~(\ref{hami1})
since the spin splitting~\cite{split} is not visible in the energy levels
shown in Fig.~\ref{exp_spectrum} for the magnetic fields applied.

The confining potential $V_{\rm conf}({\mathbf r})$ is 
expected to be parabolic near the center of the dot.
However, we soften the edges of the dot by changing
the sign of the paraboloid at a certain cusp radius $r_c$, giving
\begin{equation}
V_{\rm conf}({\mathbf r})=\left\{\begin{array}{ll}
\frac{1}{2}m^*\omega^2_0 r^2, & r\leq r_c\\
m^*\omega^2_0\left[s(r-r_c)^2-r_c(\frac{r_c}{2}-r)\right], & r>r_c,
\end{array} \right.
\end{equation}
where the parameter $s$ defines the strength of the rounding term.
As shown below, the softening of the confinement is crucial
in obtaining a good agreement with the experimental energy spectrum.

\begin{figure}
\includegraphics[width=7.5cm]{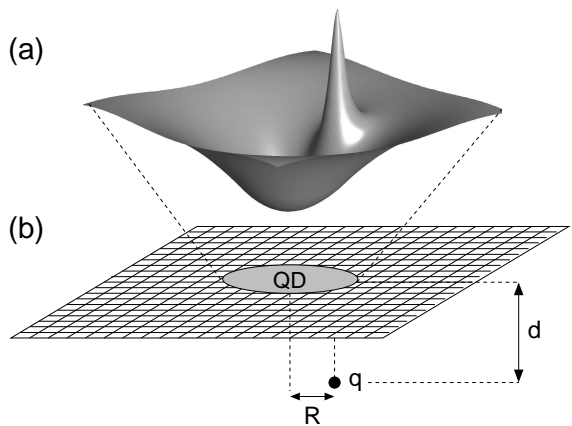}
\caption{(a) Profile of the external potential used in the simulation.
(b) Sketch of the expected configuration of the QD-impurity system.}
\label{vex}
\end{figure}

We expect the impurity to be described by a 
negatively charged particle 
located in the vicinity of the quantum well.
The impurity potential can thus be written in a Coulombic 
form as
\begin{equation}
V_{\rm imp}({\mathbf r})=\frac{|q|}{4\pi\epsilon_0\kappa\sqrt{(\mathbf r - \mathbf R)^2+d^2}},
\end{equation}
where $q$ is the (negative) charge of the impurity particle,
$\kappa$ describes the ``dielectricity'' between the impurity and 
the electrons in the QD, and $R$ and $d$ are the lateral 
and vertical distances of the impurity from the QD center, 
respectively.
Figure~\ref{vex} shows the total external confinement of the 
model system, $V_{\rm ext}=V_{\rm conf}+V_{\rm imp}$,
and a sketch of the expected configuration.

\begin{figure}
\includegraphics[width=8cm]{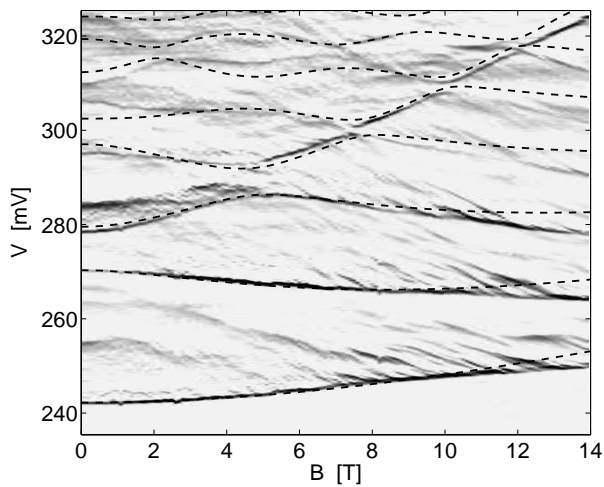}
\caption{Measured transport spectrum 
(repeated from Fig.~\ref{exp_spectrum}) of a GaAs/AlGaAs 
QD and the calculated single-electron energies 
(dashed lines) corresponding to the model potential shown
in Fig.~\ref{vex}(a).}
\label{spectra}
\end{figure}

To calculate the single-electron spectrum, we solve the
discretized eigenvalue problem $h\psi_i=\epsilon_i\psi_i$
($i=1\ldots 12$) numerically on a 2D point grid using a 
Rayleigh quotient multigrid 
method.~\cite{mika} Figure~\ref{spectra} shows the
resulting spectrum (dashed lines) compared to the experimental data 
(repeated from Fig.~\ref{exp_spectrum}). 
The energies are converted to voltages according to Eq.~(\ref{conv}), and
the model parameters are
adjusted (see the discussion below) until the agreement between 
the experiment and the model is 
as good as possible. The simulation places the avoided 
crossings between the energy levels very close to the correct positions. 
There are still considerable deviations in the 5th and 6th 
levels but, for example, the 7th level agrees almost perfectly 
through the magnetic-field regime presented. 
The differences at high fields between the experimental data 
and the simulation result from the shift of the chemical 
potential of the emitter to higher energies
with increasing magnetic field. 

In calculating the energy spectrum shown in Fig.~\ref{spectra}, 
we optimize the potential parameters corresponding to the best possible 
fit to the experimental data. The confinement is then 
defined by $\hbar\omega_0=13.8$ meV, $r_c=15.5$ nm, and $s=-0.2$, 
and the impurity parameters are given by 
$q/\kappa=-2\,e$ , $R=14.5$ nm, and $d=2$ nm.
There is naturally some uncertainty in the parameters.
However, for this particular sample we find the following characteristic 
features in the model potential: (i) The $\kappa$-reduced impurity charge
is so large that it indicates a multiple charged particle distorting the QD. 
(ii) The impurity is located very close to
the QD plane, probably lying in the 10 nm thick GaAs layer 
(see Fig.~\ref{sample}). (iii) The confining potential is
approximately three times larger than the values typically 
used ($3\ldots5$ meV) for modeling parabolic QD's.
This is due to the grown-induced formation of the QD
in the absence of gates around the sample. (iv) For the same reason,
the confinement becomes softer toward the edges of the dot. Hence, 
the rounding at $r\geq{r_c}$ in $V_{\rm conf}$ is required to 
compress the highest states in agreement with the experimental 
spectrum.

\begin{figure}
\includegraphics[width=8cm]{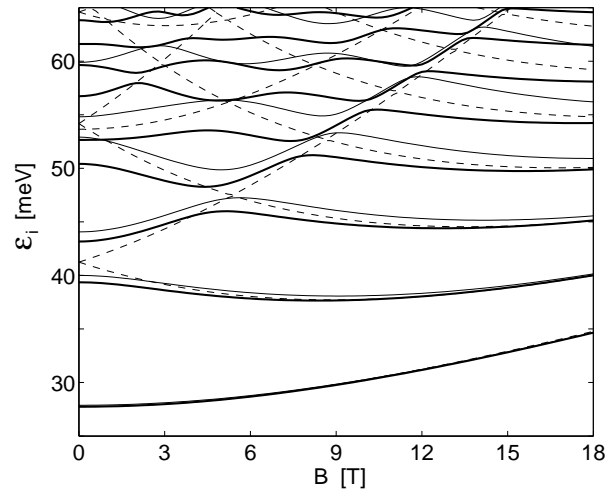}
\caption{Lowest noninteracting single-electron 
eigenenergies calculated for an impurity-containing QD 
with (thick lines) and without (thin lines) the rounding
of the edges. The dashed lines show 
the corresponding eigenenergies (lifted by +14 
meV for clarity) 
for a clean dot.}
\label{single}
\end{figure}

To clarify the sensitivity of the single-electron spectrum
on the shape of the model potential, we compare
in Fig.~\ref{single}
the eigenenergies given by the chosen model (thick lines) to those 
of a model QD without the rounded edges (thin lines) and to those of a 
corresponding clean model dot without $V_{\rm imp}$ but with the rounding term 
(dashed lines). The level repulsion is clearly induced by the 
Coulombic impurity that breaks the circular symmetry of the QD 
and mixes the FD 
levels of different orbital quantum numbers. 
A statistical analysis of the 
energy-level spacings would enlighten the quantum 
chaotic properties~\cite{stockmann} 
of the system but it is not included in this study.
In the high-field limit, however, the system becomes 
integrable and the eigenstates condense into Landau levels.

The rounding term in $V_{\rm conf}$ has the strongest 
influence on the levels with the highest angular 
momenta, and the cusp at $r_c$ induces also a 
weak decoupling of the degeneracies at $B=0$ T.
We remark that the eigenenergies for the clean case are
lifted in Fig.~\ref{single} by $14$ meV for clarity. 
The Coulombic impurity in the vicinity of the QD
thus has a strong effect on the eigenenergies.
This tendency is also apparent in the many-electron
properties studied below.

\section{Many-electron properties} \label{sec4}

Next we study situations that the quantum dot 
described by the best fitting parameters above
contains up to six interacting electrons.
Even if the many-electron case has not yet been 
experimentally realized for this 
particular QD, we find it important to predict how the 
increasing electron number changes the effects 
of the impurity on the ground-state properties. 

The problem is now described 
by the $N$-electron Hamiltonian
\begin{equation}
H=\sum^N_{i=1}\left[h_i+g^*\mu_BBs_{z,i}\right]+
+\sum^N_{i<j}\frac{e^2}{4\pi\epsilon_0\epsilon|{\mathbf r}_i-{\mathbf r}_j|},
\label{hami}
\end{equation}
where the single-electron part $h_i$ is given by Eq.~(\ref{hami1})
and the Zeeman energy is taken into account with 
$g^*=-0.22$ for the effective gyromagnetic ratio in GaAs.
This has been measured to be a realistic value for a similar 
system.~\cite{unpub} We have $\epsilon=12.7$ for the dielectric
constant in the Coulomb interaction between the electrons.

We apply the SDFT according to the self-consistent 
Kohn-Sham (KS) scheme to obtain the total energies and
spin densities of the system.
The local spin-density approximation used for the
exchange-correlation energy is based on the
magnetic-field-independent formulation by
Attaccalite {\em et al.} \cite{attaccalite}
According to our experience, it is the most accurate parametrization
for finite two-dimensional electron systems
in the zero-field limit.~\cite{lsda}
We solve the discretized KS equations in real space
without implicit symmetry restrictions, which
allows us a total freedom in shaping the
external potential. The numerical process 
of solving the effective single-electron 
Schr\"odinger equation is
accelerated with the Rayleigh quotient multigrid
method.~\cite{mika}

Our earlier calculations for rectangular~\cite{zerorecta} 
and parabolic~\cite{lsda} QD's show that 
our SDFT scheme produces energies in a good
accordance with the quantum Monte Carlo results.
We have also noticed that the current-spin-density-functional 
theory (CSDFT) does not represent a qualitative improvement 
over the SDFT in small quantum-dot systems. 
A detailed comparison between the SDFT and CSDFT for a 
six-electron QD can be found in Ref.~\onlinecite{lsda}.

\subsection{Energies}

\begin{figure}
\includegraphics[width=7.5cm]{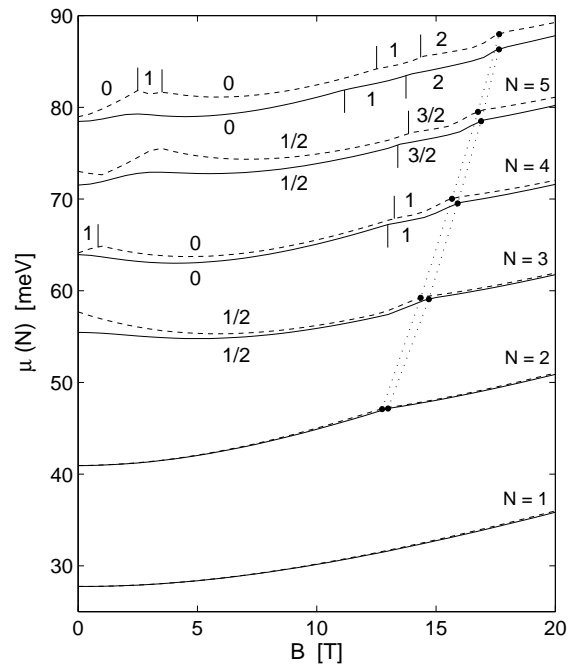}
\caption{Calculated chemical potentials
for a clean (dashed lines, lifted by $+14$ meV for clarity) 
and impurity-containing (solid lines) quantum dot up to six 
electrons. The dotted lines denote
the borders for the full spin polarization. The other spin-state
domains are also marked above and below the curves 
corresponding to the clean and 
impurity-containing cases, respectively.}
\label{chem}
\end{figure}

Figure~\ref{chem} shows the chemial potentials
$\mu(N)=E(N)-E(N-1)$, calculated in clean (dashed lines) and 
impurity-containing (solid lines) QD's up to six electrons. 
Due to the spin-degeneracy, the chemical potentials shift generally 
in adjacent pairs with increasing $B$ corresponding to the well-known 
even-odd effect of experimental current peaks.~\cite{tarucha}
In the clean case, however, occasional pairing of 
$\mu(N+1)$ and $\mu(N-1)$ is clearly observable, e.g., with
$N=3$ and $5$, and with $N=4$ and $6$ at $B\lesssim 1$ T.
This is caused by Hund's rule, i.e., near a degenerate point it is 
energetically favorable to have parallel spins between the  
electrons due to the exchange interaction. 
This leads to partial spin polarization ($S=0\rightarrow 1$) in the 
clean dot with $N=4$ and $N=6$ at $B\leq 0.84$ T and 
$2.5\,{\rm T}\leq B\leq 3.5\,{\rm T}$, respectively. 
In the impurity-containing dot those states are missing due to
the avoided level crossings in the single-electron spectrum 
(see Fig.~\ref{single}). Generally, the impurity smoothes
the behavior of the chemical potential, reflecting the flattening 
of the single-electron energies studied above. This is in agreement 
with the results by G\"u\c{c}l\"u~{\em et al.}~\cite{gycly} for
randomly distributed impurities.

As seen in Fig.~\ref{chem} the impurity does not
affect considerably the onset for the full spin polarization
(dotted lines). As $N$ increases, however, this point
in the clean QD shifts to higher $B$ than in 
the impurity-containing dot. Simultaneously, the $\mu$ spacings
decrease relatively more rapidly in the latter case than in
the clean dot. The reason is the fact that the impurity 
pushes the electrons to the soft-confinement region near 
the edges, whereas the clean dot represents a more compact state
with higher addition energies. 

\subsection{Angular momentum}

In the description above, the impurity is kept 
in a fixed position at $R=14.5$ nm laterally from 
the dot center. Next, we consider changes in the
total angular momentum as $R$ is varied 
(while $d$ is kept constant) in magnetic fields 
corresponding to
the MDD state. In a circularly symmetric QD, the MDD
has the occupation on the single-particle states with 
angular momentum eigenvalues $l=0,-1,...,-N+1$, giving 
$L_z=-\frac{1}{2}N(N-1)$ for the total angular 
momentum.~\cite{yang}
In the case of a broken circular symmetry, the onset
for the MDD can be defined from the last cusp in the 
chemical potential indicating 
the full spin polarization (see Fig.~\ref{chem}).
This is consistent with the experimental 
identification of the MDD phase by measuring the
magnetic field evolution of the Coulomb blockade 
peaks.~\cite{oosterkamp} Since the angular momentum
$l$ is not a good quantum number for a non-circular
confinement, we compute $L_z$ in the following 
as a sum of the expectation
values for the single-electron angular momentum operator, 
$\hat{l}_z=-i\hbar[x(\partial/\partial{y})-y(\partial/\partial{x})]$. 
The summation is over all the occupied KS states.

\begin{figure}
\includegraphics[width=8cm]{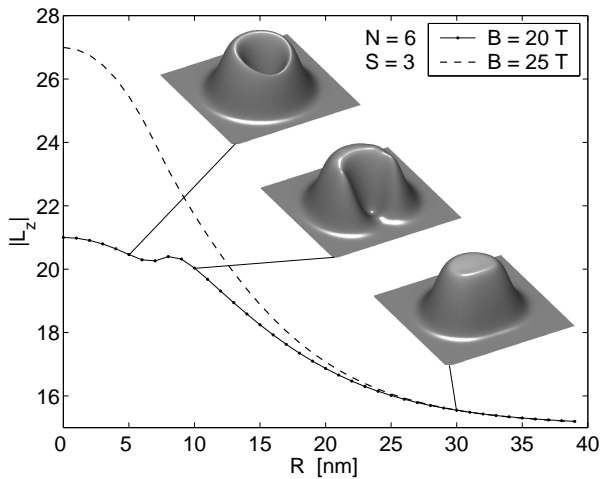}
\caption{Total angular momentum and electron density distributions
for a six-electron QD as a function of the impurity
distance from the dot center. The solid and dashed curves correspond
to the magnetic fields of $20$ and $25$ T, respectively.}
\label{lzdens}
\end{figure}

In Fig.~\ref{lzdens}
we present how the six-electron MDD at $B=20$ T 
is affected by the impurity 
shifted through the whole QD region. 
Except for the kink at $R\sim 8$ nm, corresponding
to the radius at which the impurity penetrates the
edge of the QD, $|L_z|$ increases smoothly from $15$ to $21$, i.e., from the
MDD to a quantum ring with $l=-1,...,-6$ states occupied. 
The stability
is also visualized in the density insets of Fig.~\ref{lzdens} 
corresponding to different impurity locations. 
The dashed line shows the equivalent evolution at $B=25$ T, still 
corresponding to the MDD region in the clean dot. As $R\rightarrow 0$, 
however, we find a transition to the next quantum-ring 
state with $l=-2,...,-7$ states occupied. 
The rounded edges of the QD clearly make the ring-like states 
sensitive to the magnetic field. In contrast, the MDD remains as a 
rather compact state spread across the highly-confined 
central region as the magnetic field is varied.

\subsection{Magnetization}

\begin{figure}
\includegraphics[width=8cm]{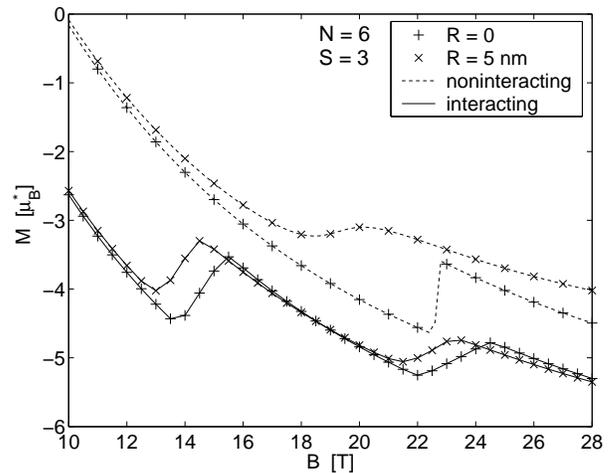}
\caption{Magnetization of the (fully-polarized) state
of a six-electron quantum dot at the impurity distances 
$R=0$ (plus markers) and $5$ nm (cross markers)
with (solid curves) and without (dashed curves) 
the electron-electron interactions.
The effective Bohr magneton $\mu_B^*=e\hbar/2m^*$.}
\label{magnet}
\end{figure}

Finally, we investigate the tendency of the many-electron
structure to screen the external impurity.
For that purpose, we consider the total magnetization, 
defined at zero temperature as 
$M=-{\partial E_{\rm tot}}/{\partial B}$.
In Fig.~\ref{magnet} we plot $M$
as a function of $B$ for six-electron, 
fully-polarized QD's in both noninteracting and interacting 
cases and with the impurity locations $R=0$ and $5$ nm
laterally from the QD center.
The noninteracting energy is computed directly as a
sum of the six lowest single-electron energies.
As the impurity is shifted off the dot center,
the lowest single-electron states decouple
(see Fig.~1 in Ref.~\onlinecite{halonen}), leading to 
a considerably smaller $|M|$ and disappearance of the 
transition at $B\sim 23$ T in the noninteracting systems.
On the other hand, the interacting systems show frequent oscillations
due to the entanglement of the energy levels,
i.e., periodic changes in the fully-polarized ground state.~\cite{recta}
The QD's with different impurity locations 
behave similarly, indicating that the circular
properties are preserved also when the impurity is 
off-center. This verifies the screening of the impurity by the 
electrons.
The importance of screening is evident also from the work
by Halonen~{\em et al.}~\cite{halonen} who
considered impurity-containing QD's with two and three electrons. 
The same tendency was also found in the magnetization 
of the two-electron square QD's studied by Sheng and 
Zu.~\cite{sheng} In our forthcoming studies, we will 
examine the high magnetic field regime as a function
of the impurity parameters in more detail. 

\section{Summary} \label{sec5}

We have reported a transport measurement
of a vertical quantum dot, giving single-electron
energies strongly deviating from the Fock-Darwin spectrum.
We have found a realistic model for the sample dot by
including a Coulombic impurity in the external
parabolic potential that has soft edges. 
The parameters obtained for the model potential
indicate a strong confinement resulting from the
grown-induced formation of the QD, and
a repulsive, multiple charged impurity particle 
shifted both laterally and vertically from the QD center.
Using the model and the spin-density-functional
theory, we have studied the energetics 
and structural properties of the corresponding 
many-electron problem.
The impurity evens out the state alternation as a function
of the magnetic field by lifting the degeneracies,
but it does not affect remarkably the onset of the full
spin polarization. The stability of both the maximum-density
droplet and the magnetization verify that the 
many-electron structure reduces the distortion 
apparent in the single-electron properties.
The future experiments may illustrate
the damping of the irregular behavior in actual quantum-dot 
devices as the number of electrons increases.

\begin{acknowledgments}

This research was supported by the Academy of Finland through
its Centers of Excellence program (2000-2005).
E.R. is also grateful for Magnus Ehrnrooth Foundation for
financial support and A. Harju and A. D. G\"u\c{c}l\"u for useful discussions.
The group in Hannover acknowledges sample growth by V. Avrutin and 
A. Waag and financial support by BMBF.

\end{acknowledgments}


\begin{thebibliography}{99}

\bibitem{revmod} For a review, see, e.g., L. P. Kouwenhoven, D. G. Austing,
and S. Tarucha, Rep. Prog. Phys. {\bf 64}, 701 (2001); S. M. Reimann and 
M. Manninen, Rev. Mod. Phys. {\bf 74}, 1283 (2002).

\bibitem{fd} V. Fock, Z. Phys. {\bf 47}, 466 (1928); 
C. G. Darwin, Proc. Cambridge Philos. Soc. {\bf 27}, 86 (1930).

\bibitem{halonen} V. Halonen, P. Hyv\"onen, P. Pietil\"ainen, 
and T. Chakraborty, Phys. Rev. B {\bf 53}, 6971 (1996). 

\bibitem{mceuen} P. L. McEuen, E. B. Foxman, U. Meirav, 
M. A. Kastner, Y. Meir, N. S. Wingreen, and S. J. Wind, 
Phys. Rev. Lett. {\bf 66}, 1926 (1991);
J. Weis, R. J. Haug, K. v. Klitzing, and K. Ploog, 
Phys. Rev. B {\bf 46}, 12837 (1992).

\bibitem{vertical} T. Schmidt, M. Tewordt, R. H. Blick, R. J. Haug, D.
Pfannkuche, K. v. Klitzing, A. F\"orster, and H. L\"uth, 
Phys. Rev. B {\bf 51}, 5570 (1995);
J. K\"onemann, D. K. Maude, V. Avrutin, A. Waag, and R. J. Haug,
Physica E {\bf 22}, 434 (2004).

\bibitem{tarucha} S. Tarucha, D. G. Austing, T. Honda,
R. J. van der Hage, and L. P. Kouwenhoven,
Phys. Rev. Lett. {\bf 77}, 3613 (1996).

\bibitem{ashoori} R. C. Ashoori, H. L. Stormer, J. S. Weiner,
L. N. Pfeiffer, S. J. Pearton, K. W. Baldwin, and K. W. West,
Phys. Rev. Lett. {\bf 68}, 3088 (1992).

\bibitem{wan} Y. Wan, G. Ortiz, and P. Phillips, 
Phys. Rev. B {\bf 55}, 5313 (1997). 

\bibitem{lee} E. Lee, A. Puzder, M. Y. Chou, T. Uzer, and D. Farrelly,
Phys. Rev. B {\bf 57}, 12281 (1998). 

\bibitem{gycly} A. D. G\"u\c{c}l\"u, J. S. Wang, and H. Guo, 
Phys. Rev. B {\bf 68}, 035304 (2003).

\bibitem{hirose} K. Hirose and N. S. Wingreen, 
Phys. Rev. B {\bf 65}, 193305 (2002). 

\bibitem{ahn} K.-H. Ahn, K. Richter, and In-Ho Lee, 
Phys. Rev. Lett. {\bf 83}, 4144 (1999). 

\bibitem{recta} E. R\"as\"anen, A. Harju, M. J. Puska, and R. M. Nieminen,
Phys. Rev. B {\bf 69}, 165309 (2004).

\bibitem{split} J. K\"onemann, P. K\"onig, and R. J. Haug,
Physica E {\bf 13}, 675 (2002).

\bibitem{mika} M. Heiskanen, T. Torsti, M. J. Puska, and R. M. Nieminen,
Phys. Rev. B {\bf 63}, 245106 (2001).

\bibitem{stockmann} H.-J. Stockmann, {\em Quantum Chaos: An Introduction}
(Cambridge University Press, Cambridge, 2000).

\bibitem{unpub} J. K\"onemann {\em et al.}, to be published.

\bibitem{attaccalite} C. Attaccalite, S. Moroni, P. Gori-Giorgi, and
G. B. Bachelet, Phys. Rev. Lett. {\bf 88}, 256601 (2002);
{\bf 91}, 109902(E) (2003).

\bibitem{lsda} H. Saarikoski, E. R\"as\"anen, S. Siljam\"aki, A. Harju,
M. J. Puska, and R. M. Nieminen, Phys. Rev. B {\bf 67}, 205327 (2003).

\bibitem{zerorecta} E. R\"as\"anen, H. Saarikoski, V. N. Stavrou, A. Harju,
M. J. Puska, and R. M. Nieminen, Phys. Rev. B {\bf 67}, 235307 (2003).

\bibitem{yang} S.-R. Eric Yang and A. H. MacDonald,
Phys. Rev. B {\bf 66}, 041304(R) (2002), and references therein.

\bibitem{oosterkamp} T. H. Oosterkamp, J. W. Janssen,
L. P. Kouwenhoven, D. G. Austing, T. Honda, and S. Tarucha,
Phys. Rev. Lett. {\bf 82}, 2931 (1999).

\bibitem{sheng} W. Sheng and H. Xu, Physica B {\bf 256-258}, 152 (1998).

\end{thebibliography}
\end{document}